\documentclass[12pt]{article}
\begin{document}

\title{Comments on ``An Exact, Three-Dimensional, Time-Dependent Wave
Solution in Local Keplerian Flow'' by Balbus and Hawley
(astro-ph/0608429)}

\author{G. D. Chagelishvili and A. G. Tevzadze \\
E. Kharadze Abastumani National Astrophysical Observatory, \\
2a Al. Kazbegi Ave. Tbilisi 0160, Georgia, \\
E-mail: g.chagelishvili@astro-ge.org}

\maketitle

\begin{abstract}
We analyze the scepticism on the hydrodynamic turbulence in
Keplerian astrophysical disks expressed in Balbus and Hawley 2006
and show the failure of arguments of the paper.
\end{abstract}

Recently Balbus and Hawley 2006 (BH06) have shown time-dependent
three-dimensional nonlinear wave solution when considering a local
Keplerian flow. Authors have demonstrated that the linear wave
solution derived in Johnson and Gammie (2005) is actually also valid
nonlinearly when considering a single plane wave. This solution can
be used as a test for the dissipative properties of the numerical
codes. The key message of BH06 is given in the abstract: \emph{"The
existence of this solution argues against transient amplification as
a route to turbulence in unmagnetized disks"}. However, this message
does not correctly describes the situation in unmagnetized
astrophysical disks. In the latter case the determinative factor of
local dynamics is vertical stratification of the flow, which is
neglected by BH06: the stratification originates the modes of
perturbations that are able to withstand stabilizing influence of
Coriolis forces. Dynamics of local 3D perturbation modes in stably
stratified Keplerian flow is studied in detail by Tevzadze at el.
2003 (T03).

BH06 have omitted the key factors that provide the fundamental
source of energy of the turbulence: the combined action of vertical
gravity and Coriolis forces in 3D case results the conservation of
the potential vorticity (see Eq. 11 of T03) that indicates the
existence of a vortex/apperiodic mode in the perturbation spectrum
of the system. This vortex mode is the primary in the transient
amplification phenomenon. In the absence of any one of these
(vertical gravity or Coriolis) forces vortex mode degenerates into
the trivial solution of the system -- it disappears. Thus, in the
unstratified Kepler flow considered in BH06 the vortex mode is
absent, i.e.,  the main extractor of the shear flow energy is
missing. Hence, it is incorrect to judge ``against transient
amplification as a route to turbulence in unmagnetized disks'' based
on the model that neglects the main energy source of the turbulence.

In three-dimensional hydrodynamic disks with vertical gravity the
spiral wave mode considered in BH06 corresponds to the
density-spiral wave that is internal gravity waves modified by the
disk rotation. This wave does not extracts the energy of
differential rotation efficiently, but is linearly coupled with the
vortex mode. This coupling indicates the importance of the
density-spiral waves along with the vortices in the overall dynamics
of the system. The linear dynamics of small scale perturbations can
be analyzed by a single spatial Fourier harmonic (see Eq. 9a of T03
and Eq. 25 of BH06, where spatial Fourier harmonics are referred to
as plane waves): a leading Fourier harmonic of three-dimensional
vortex mode gains background shear flow energy and are transiently
amplified by the several orders of magnitude. Reaching the point of
maximal amplification and switching to the trailing, it gives rise
to the corresponding harmonic of the density-spiral wave due to the
shear flow induced coupling. The generated wave maintains the energy
of perturbations. In fact, overall energetic dynamics is similar to
that occurred in the plane parallel constant shear flow (see Fig. 2
of T03). Hence, it is concluded that the linear dynamics in
vertically stratified three-dimensional hydrodynamic Keplerian disks
matches requirements of the bypass concept developed for the
plane-parallel flows. (A simple sketch of the bypass scenario
revealing its essence can be found in Chagelishvili et al. 2003.)

BH06 criticizes the hydrodynamic disk turbulence model also due to
the fact that: \emph{"linear theory and nonlinear theory are the
same for a finite amplitude plane wave. ... the neglected nonlinear
wave term leads not to a viable free energy source, but to mutual
wave-wave interactions."} However, it is common knowledge in
hydrodynamic community that the same situation is in spectrally
stable plane-parallel shear flows that exhibit self-sustained
turbulence. And it is precisely the bypass concept that was designed
in the last decade of the 20th century in hydrodynamic community to
describe onset of turbulence in such flows. There is no need in
nonlinear energy sources in the bypass concept: nonlinear
interactions (i.e. interactions of plane waves with different
wavenumbers) are responsible only for redistribution of the plane
waves in the wavenumber space. Thus, after linear transient growth
nonlinear interactions repopulate exhausted leading plane waves and,
in doing so, close the feedback loop of bypass transition to
turbulence.

Finally, we would like to stress the confusion of \emph{nonlinear
instability} with \emph{transient linear growth} occurred not only
in the commented paper. Smooth shear flows (without inflection point
in the velocity profile) are spectrally stable, meaning that
exponentially growing solutions are absent. However, it is well
known that finite amplitude perturbations sometimes lead to the
transition from laminar to turbulent state at moderate, less then
critical Reynolds number. At the initial stages of the research the
concept of nonlinear instability was developed in hydrodynamics to
address this issue (see Chagelishvili et al. 2003 for references and
detail description). Thus, until about a relatively recent years,
the predominant view of this laminar-turbulent transition was
focused around the slow linear amplification of exponentially
growing perturbations (the familiar T-S waves), which modify the
flow profile and thereby allow a secondary instability, further
nonlinearity and finally a breakdown to turbulence. According to
this concept the perturbations (and the turbulent state itself) are
energetically sustained by nonlinear processes. This concept of
\emph{nonlinear instability} has been borrowed by astrophysicists,
who still use it to explain turbulent processes in smooth
astrophysical flows, where no spectrally unstable solution is known
and in particular in Keplerian disk flows. However, next step in the
understanding of the hydrodynamic turbulence have resulted the
conceptual change of the turbulence transition scenario during the
last decade of the 20th century. Another viewpoint -- referred as
the bypass transition -- has been developed in the hydrodynamic
community on the understanding of the onset of turbulence in
spectrally stable shear flows. Although the bypass transition
scenario involves nonlinear interactions -- which intervene once the
perturbations have reached finite amplitude -- the dominant
mechanism leading to these large amplitudes appears to be linear.
This concept of the onset of turbulence is based on the linear
transient growth of vortex mode (aperiodic) perturbations and is
principally different from the concept of nonlinear instability.

{~\\ \large \bf References \\} {\small
~\\
Balbus, S. A., and Hawley, J. F., 2006 (BH06), {\it
``Three-Dimensional, Time-Dependent Wave Solution in Local Keplerian
Flow''}, astro-ph/0608429 \\
\\
Chagelishvili, G. D., Zahn, J.-P., Tevzadze, A. G.,  \& Lominadze,
J. G. 2003, {\it ``On Hydrodynamic Shear Turbulence in Keplerian
Disks: From Transient Growth to Bypass Transition``}, A\&A, {\bf
402}, 401 \\
\\
Johnson, B. M., and Gammie, C. F. 2005, {\it ``Linear Theory of
Thin, Radially Stratified Disks''}, ApJ, {\bf 626}, 978 \\
\\
Tevzadze, A. G., Chagelishvili, G. D., Zahn, J.-P., Chanishvili, R.
G. \& Lominadze, J. G. 2003 (T03), {\it ``On Hydrodynamic Shear
Turbulence in Keplerian Disks: Transient Growth of 3D Small Scale
Perturbations''}, A\&A, {\bf 408}, 779

\end{document}